\documentclass[aps,prd,showkeys,superscriptaddress,twocolumn]{revtex4}
\usepackage{epsfig}
\usepackage{graphicx}
\usepackage{amsmath,amssymb,amsfonts,latexsym}
\usepackage{graphicx}

\usepackage{epsfig,amssymb,amsfonts,verbatim}

\usepackage{amsmath}
\usepackage{latexsym}
\usepackage{amsfonts}
\usepackage{amssymb}
\usepackage{color}

\def\bfl{\begin{flushleft}}
\def\efl{\end{flushleft}}
\def\bfr{\begin{flushright}}
\def\efr{\end{flushright}}
\def\bc{\begin{center}}
\def\ec{\end{center}}

\def\ba{\begin{eqnarray}}
\def\ea{\end{eqnarray}}
\def\baa#1{\begin{array}{#1}}
\def\eaa{\end{array}}
\def\bw{\begin{widetext}}
\def\ew{\end{widetext}}

\def\text#1{\mbox{#1}}

\begin{document}

%\preprint{APS/123-QED}

\title{Diffusivity of adatoms on plasma-exposed surfaces determined from the ionization energy approximation and ionic polarizability}

\author{Andrew Das Arulsamy}
\email{andrew@physics.usyd.edu.au}
\affiliation{School of Physics, The
University of Sydney, Sydney, New South Wales 2006, Australia}

\author{Kostya (Ken) Ostrikov}
\affiliation{CSIRO Materials Science and Engineering, P.O. Box 218, Lindfield NSW 2070, Australia}
\affiliation{School of Physics, The University of Sydney, Sydney, New South Wales 2006, Australia}

\date{\today}
%\narrowtext
%\small

\begin{abstract}
Microscopic surface diffusivity theory based on atomic ionization energy concept is developed to explain the variations of the atomic and displacement polarizations with respect to the surface diffusion activation energy of adatoms in the process of self-assembly of quantum dots on plasma-exposed surfaces. These polarizations are derived classically, while the atomic polarization is quantized to obtain the microscopic atomic polarizability. The surface diffusivity equation is derived as a function of the ionization energy. The results of this work can be used to fine-tune the delivery rates of different adatoms onto nanostructure growth surfaces and optimize the low-temperature plasma based nanoscale synthesis processes. 
\end{abstract}

%\pacs{73.21.La; 73.21.-b; 68.65.-k}

\maketitle

\section{Introduction}

Self-organization and assembly of quantum dots (QDs) on surfaces exposed to low-temperature plasmas play crucial roles in the growth of QD arrays with uniform positioning and size distributions~\cite{osk,os1,os2,gor,kei}. In addition, self-organization is the only way to develop arrays of QDs in the size range below the capability of currently existing ``top-down" nanofabrication tools~\cite{os2,uros,mele,dren}. The significance of such uniformity is to avoid the inhomogeneous broadening in quantized energy levels that may adversely affect the atomic-like properties arising from individual QDs~\cite{cunt,raja}. Plasma-based nanofabrication techniques possess pronounced advantages to produce uniform QDs in terms of distributions of QD sizes, as well as the distances between QDs~\cite{ho}. This uniformity is extremely important for applications in different fields, for example, in DNA-single electron transistors (nanoelectronics)~\cite{rak}, in photodynamic and radiation therapies (medicine)~\cite{pet}, in nanophotonics (quantum information)~\cite{jia,eng,kumar,singh} and nano-optics~\cite{liang,jang,nemec,hsiang}. 

Self-organization in large QD arrays requires the consideration of surface diffusion in the presence of microscopic electric fields~\cite{os1}. One can take this effect into account by means of electric field gradient (which depends on QD radii and the mutual positioning on the array) and the electric dipole moment of adatoms. For example, it was calculated that the energy taken by an adatom in one jump across one lattice spacing is $W_e = |\partial \vec{\textbf{E}}_{\rm{app}}/\partial r|\lambda_a p$, where $\vec{\textbf{E}}_{\rm{app}}$, $\lambda_a$ and $p$ denote the applied electric field (microscopic component in the vicinity of the developing nano-pattern), lattice parameter and the electric dipole moment, respectively~\cite{os1}. Therefore, larger electric field gradients (which are produced by smaller QDs) may increase the surface diffusion rates through lowering the surface diffusion activation energy. This in turn leads to higher rates of adatoms leaving the smaller QDs; these adatoms become available to nucleate new QDs~\cite{os1}. 

The proportionality between the diffusion coefficient and the temperature-dependent diffusion are established based on the Nernst-Einstein relationship~\cite{sharma}. As a consequence, one can surmise here that for a given (i) QDs (material), (ii) applied electric field, (iii) substrate material and (iv) surface temperature, one can control the uniformity of the QD arrays. In this work, another step forward is taken to study how the plasma influx needs to be varied for different QDs on different substrates for given  applied electric field and temperature ($T$). In order to achieve this objective, we microscopically derive the atomic polarizability and use it to calculate the displacement polarizability of adatoms with respect to different elemental composition in QDs. Reliable knowledge of these characteristics is indispensable since surface diffusion of adatoms is the main driving force for self-organization of QDs in microscopic patterns~\cite{os1,os2,ars,john,deg}. Therefore, we will employ the recently proposed approach of controlling the surface diffusion of adatoms via the applied electric field, which is important to determine the rates of QD formation~\cite{os1,os2}. Subsequently, the ionization energy theory that considers the effect of different elemental composition~\cite{arul1,arul2,arul3,arul4} will be used to predict the electric-field dependent surface diffusivity of adatoms on the plasma-exposed surfaces. This coupled approach can be used to derive the above polarizabilities and to calculate the surface diffusivity of different adatoms in the presence of applied electric field and at specific process temperatures. We will also discuss how to fine-tune the delivery rates of different species from the plasma to synthesize multi-element QDs using the calculated rates of the surface diffusion. Si$_{1-x}$C$_x$ QDs on a Si substrate will be used as an example throughout in this work due to their promising applications in photovoltaic devices~\cite{varn,martin,ferre}. 

%We organize the paper in the following order. The dressed phonon frequency and its relation with the many-body Hamiltonian are introduced with respect to the ionization energy theory. Next, both semi-classical and quantized derivations of the atomic polarizability as a function of the ionization energy are given. Subsequently, displacement (ionic) polarizability for deformable ions are derived and discussed due to its influence on QDs growth. Before conclude the paper, comprehensive discussion on controlling the applied electric field and temperature are given based on the polarizability functions derived in the earlier sections.

\section{Theory of the polarizability applied to quantum dots}

Here, both atomic (electronic) and displacement (ionic) polarizabilities are derived as functions of the ionization energy, and subsequently, their relation to the surface diffusion coefficient for the adatoms are discussed. These derivations will be the backbone for the latter comprehensive discussion on controlling the nanoscale synthesis process parameters, such as the influx ratios of precursor species.  

\subsection{Many-body Hamiltonian}

To calculate the polarizability and the adatom diffusion rates on a given substrate, we start from the dressed phonon frequency, which is given by~\cite{arul1,arul3}

\begin {eqnarray}
\omega(\xi,\textbf{k}) = \frac{k\Omega_p}{K_s}\exp\bigg[\frac{1}{2}\lambda(\xi - E_F^0)\bigg], \label{eq:1}
\end {eqnarray}

where, $\Omega_p$ is the ionic plasma frequency in a solid (long-wavelength density oscillations) and $\textsl{K}_s^2$ = $3n_0e^2/2\epsilon_0 E_F^0$. Here, $n_0$ and $E_F^0$ are the respective carrier density and the Fermi level or the highest occupied energy level for QDs, both at $T$ = 0. Furthermore, $k$ and $\textsl{K}_s$ are the wavenumber and the Thomas-Fermi wavenumber, respectively~\cite{arul1,arul3}. The exponential term in Eq.~(\ref{eq:1}) is a function of the ionization energy, $\xi$ has been derived from the carrier density ($n$) equation as given below

\begin {eqnarray}
n = \int^{\infty}_{0} f_e(E)N_e(E)dE, \label{eq:2}
\end {eqnarray}
   
where $f_e$ and $N_e$ are the ionization energy-based electronic probability function and the density of states, respectively. Hence, it is easy to notice here that this exponential term comes from $f_e$, which is given by~\cite{arul2,arul3} 

\begin{eqnarray}
f_e(E_0,\xi) = \frac{1}{e^{\lambda[\left(E_0 + \xi
\right) - E_F^{(0)}]}+1}, \label{eq:3}
\end{eqnarray} 

where $\lambda = (12\pi\epsilon_0/e^2)a_B$, $a_B$ is the Bohr radius of atomic hydrogen, $e$ and $\epsilon_0$ are the electronic charge and the permittivity of space, respectively~\cite{arul3}. Note that $\lambda$ can also be equal to $1/(k_BT)$, depending on applications~\cite{arul2}, where $k_B$ and $T$ denote the Boltzmann constant and temperature, respectively. For holes, one simply replaces the $+$ sign in $E_0 + \xi$ (see Eq.~(\ref{eq:3})) with the sign, $-$. Next, the term $E_0 + \xi$ (total energy) in the probability function originated from the derivation of Eq.~(\ref{eq:3})~\cite{arul1,arul2}. Finally, the total energy has originated from the many-body Hamiltonian~\cite{arul3,arul4}, 

\begin {eqnarray}
\hat{H}\varphi = (E_0 \pm \xi)\varphi. \label{eq:4}
\end {eqnarray}     

The relevant proofs for Eq.~(\ref{eq:4}) and its relation with many-electron atomic Hamiltonian are given in Ref.~\cite{arul1,arul2}. Having found the source of the exponential term, one can now work on the classical derivation of the atomic polarizability. In other words, using Eq.~(\ref{eq:1}), we can modify the classical model of the atomic polarizability. After that, it is straightforward to transform the classical version to a quantum mechanical one, where the exponential term will stay intact.  

\subsection{Atomic polarizability: Semiclassical}

\begin{figure}[tb!]
\begin{center}
%\begin{figure}[tb!]
%\centering
\scalebox{0.25}{\includegraphics{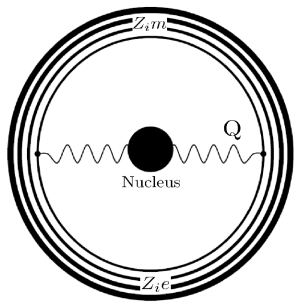}}
\caption{Semiclassical model for the atomic polarizability with discreet energy levels, and with $\xi$ as the energy level difference or the ionization energy. Recall that $\xi$ changes with different atoms and so does the interaction potential constant or the so-called spring constant, $Q$. The discreet energy levels are for \textit{non-free electrons} with total electronic charge $Z_ie$ and mass $Z_im$.}
\label{fig:1}
\end{center}
%\vspace{0.7cm}
\end{figure}

The original work on atomic polarizability entirely based on classical physics was carried out by Lorentz~\cite{lor}. We will use a similar procedure, but with quantum mechanical properties incorporated into the interaction potential constant. Before going deeper into the polarizability calculations, it is important to recall the implications of Eq.~(\ref{eq:1}). As already shown in Ref.~\cite{arul1}, the exponential term in Eq.~(\ref{eq:1}) also implies that the harmonic potential energy ($\phi$) for an atom is given by (see Fig.~\ref{fig:1})

\begin {eqnarray}
\phi(x) = \frac{1}{2}Z_imx^2\omega_0^2\exp\bigg[\lambda(\xi - E_F^0)\bigg], \label{eq:5}
\end {eqnarray}     

where, $Z_i$ and $m$ are the $i^{\rm{th}}$ atomic number and electron mass, respectively, while $\omega_0$ denotes the frequency of the vibration of the electronic shell (attached to the static nucleus via the spring as indicated in Fig.~\ref{fig:1}). The $Z_ie$ here represents the total charge of screened electrons with discreet energy levels, unlike the usual free electrons as shown in Fig.~\ref{fig:1}. Therefore, the interaction potential constant, $Q$ (also known as the spring constant) can be obtained from Eq.~(\ref{eq:5}),

\begin {eqnarray}
Q = \frac{\partial^2 \phi}{\partial x^2} = Z_im\omega_0^2\exp\bigg[\lambda(\xi - E_F^0)\bigg]. \label{eq:6}
\end {eqnarray}     

%%%%%%%%%%%%%%%%%%%%%%%%%%%%%%%%%%%%%%%%%%%%%%%%%%%%%%%%%%%%%%%%%%%%%%%%%%%%%%%%%%%%%%%%%%%%%%%%%%%%%%%%%%%%%%%%%%%%%%%%%%%%%
%\begin{figure}[hbtp!]
%%%%%%%%%%%%%%%%%%%%%%%%%%%%%%%%%%%%%%%%%%%%%%%%%%%%%%%%%%%%%%%%%%%%%%%%%%%%%%%%%%%%%%%%%%%%%%%%%%%%%%%%%%%%%%%%%%%%%%%%%%%%

Now, for mathematical convenience one can also write the frequency ($\omega$) dependent, local electric field ($\vec{\textbf{E}}$) as~\cite{ash}

\begin {eqnarray}
\vec{\textbf{E}} = \vec{\textbf{E}}_0 e^{-i\omega(\xi) t}, \label{eq:7}
\end {eqnarray}     

where, $\omega(\xi) = \omega\exp[(\lambda/2)(\xi - E_F^0)]$. The local electric field is the microscopic field from the ion itself, and any variation to that ion will also be captured by this local electric field. For a cubic-like crystal, in the spherical ion approximation, this local electric field is given by~\cite{ash,arul2} $\vec{\textbf{E}} = \vec{\textbf{E}}_{\rm{mac}} + (1/3\epsilon_0)\textbf{P}$. Here, $\textbf{P}$ is the electronic polarization and $\vec{\textbf{E}}_{\rm{mac}}$ is the macroscopic electric field. In the ionization energy theory, the constraints stated above (cubic-like crystals and spherical ions) do not arise if Eq.~(\ref{eq:1}) is used. For example, the ionization energy (or the valence states) will be different for each different (i) ion, (ii) crystal structure, and (iii) defect in a given crystal, because all issues stated in (i) to (iii) will contribute differently and cause fluctuations in the ionic valence states. In addition, since the ionization energy theory is almost entirely based on the averaged probability function, the knowledge whether the electronic polarization is parallel to the electric field or not, does not arise as well. The term, $\omega(\xi)$ introduced in Eq.~(\ref{eq:7}) means that the frequency-dependent electric field can be varied accordingly for different atoms, so as to maintain the magnitude of the atomic polarizability (this introduction is purely for mathematical convenience). 

%%%%%%%%%%%%%%%%%%%%%%%%%%%%%%%%%%%%%%%%%%%%%%%%%%%%%%%%%%%%%%%%%%%%%%%%%%%%%%%%%%%%%%%%%%%%%%%%%%%%%%%%%%%%%%%%%%%%%%%%%%%%%
\begin{figure}[tb!]
\begin{center}
%\begin{figure}[tb!]
%\centering
\scalebox{0.3}{\includegraphics{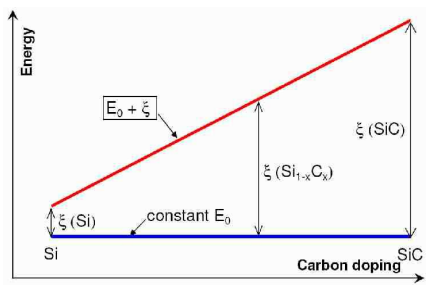}}
\caption{The evolution of the ionization energy with C substitution into Si sites (not to scale). Note here that $E_0$ is a constant and $\xi$ captures the linear relation between the energy and $\xi$ (or $x$).}
\label{fig:2}
\end{center}
%\vspace{0.7cm}
\end{figure}
%%%%%%%%%%%%%%%%%%%%%%%%%%%%%%%%%%%%%%%%%%%%%%%%%%%%%%%%%%%%%%%%%%%%%%%%%%%%%%%%%%%%%%%%%%%%%%%%%%%%%%%%%%%%%%%%%%%%%%%%%%%%

For example, if one substitutionally dopes C into Si sites to obtain Si$_{1-x}$C$_x$ QDs, then there will be a systematic change to the ionization energy ($\xi$) linearly as shown schematically in Fig.~\ref{fig:2}.  As a consequence, the frequency of the electric field can be changed accordingly with carbon content (as follows from Eq.~(\ref{eq:7})), in order to maintain the magnitude of atomic polarizability between undoped Si and doped Si$_{1-x}$C$_x$. This linear relationship stated above is of course only valid if the valence states of carbon and silicon do not change in the course of doping, which is the case for Si$_{1-x}$C$_x$. If the valence states change for each level of substitutional doping, $x$, then this will lead to fluctuations in the linear relationship. For example, defects can cause significant variations to the crystal structure and the ionic valence states. 

The electronic displacement, $\textbf{r}$ from its equilibrium, $\textbf{r}_0$ due to the local electric field is given by~\cite{ash} $\textbf{r} = \textbf{r}_0 e^{-i\omega(\xi) t}$. The equation of motion for the valence electrons around a particular ion can be written as $F = Z_im\ddot{\textbf{r}} = -Q\textbf{r} - Z_i e \vec{\textbf{E}}_0 e^{-i\omega(\xi) t}$. Eventually, one can rewrite $\textbf{r}$ to obtain 

\begin {eqnarray}
\textbf{r}_0 = -\frac{e\vec{\textbf{E}}_0}{m(\omega_0^2 - \omega^2)}\exp\bigg[\lambda(E_F^0 - \xi)\bigg]. \label{eq:10}
\end {eqnarray}  

From the definition, the induced electric dipole moment, $p$ is given by

\begin {eqnarray}
p = p_0 e^{-i\omega(\xi)t} = -Z_ie\textbf{r} = \alpha(\omega,\xi)\vec{\textbf{E}}, \label{eq:11}
\end {eqnarray} 

where $\alpha(\omega,\xi) = [Z_ie^2/m(\omega_0^2 - \omega^2)]\exp[\lambda(E_F^0 - \xi)]$, is the frequency-dependent atomic polarizability. The static polarizability can be obtained from Eq.~(\ref{eq:11}) by letting $\omega \rightarrow 0$. The exponential term derived in Eqs.~(\ref{eq:11}) is in exact form compared to the atomic polarizability used in Ref.~\cite{arul2}. 

\subsection{Atomic polarizability: Quantum mechanical}

To derive the atomic polarizability quantum mechanically, one requires the understanding of the energy spectrum in real atoms. This means that atoms have more than one natural frequency, with each frequency having its own strength factor, $f$~\cite{feyn2}. Hence, the quantum mechanical version of Eq.~(\ref{eq:11}) is given by 

\begin {eqnarray}
\alpha = \frac{Z_ie^2}{m}\exp\bigg[\lambda(E_F^0 - \xi)\bigg]\sum_j\frac{f_j}{(\omega_{0j}^2 - \omega^2)}, \label{eq:13}
\end {eqnarray}  

where the spontaneous emission (or classically known as the damping force) is ignored for simplicity. This is indeed a straightforward transformation from Eq.~(\ref{eq:11}) to Eq.~(\ref{eq:13}). The neglect of the damping factor does not affect the physics of elemental composition dependence that are being discussed here. In this transformation, the origin of the ionization energy in the exponential term ($e^{\lambda(E_F^0 - \xi)}$) is from the vibrational frequency, $\omega_0$, which is responsible for different atoms (different elemental composition) in a given compound. On the other hand, the strength factor ($f_j$) takes into account different modes of oscillations. The next issue that needs to be resolved is the displacement polarizability, which determines the rates of adatoms leaving the QDs during growth. We will address this issue in the following section.

\subsection{Displacement polarizability: Semiclassical}

Recall that the $\textbf{r}$ introduced earlier was due to electronic displacement and now the ionic (positively (Silicon)- and negatively (Carbon)-charged ions) displacement, $\textbf{u}^{\pm}$ will be studied. From Eq.~(\ref{eq:11}) we can see why Si is positively charged, which is due to $\xi_{\rm{Si^{4+}}} < \xi_{\rm{C^{4+}}}$. The dipole moment of the primitive cell (each cell contains either C or Si lattice point) is given by~\cite{ash} $P = e(\textbf{u}^{+} - \textbf{u}^{-}) = e\textbf{d}$. Note here that we do not assume that the ions are undeformable, as was done in Ref.~\cite{ash}, because the interaction potential constant ($G$), which is introduced below is identical with the one given in Eq.~(\ref{eq:6}). This newly defined potential constant takes into account the ionic deformation due to the screened core electrons via the exponential term. Using the definition of $P$, the equations of motion (of ionic Si and C) can be written as $F^+ = M^+\ddot{\textbf{u}}^+ = -G\textbf{d} + e\vec{\textbf{E}}$ and
$F^- = M^-\ddot{\textbf{u}}^- = -G(-\textbf{d}) - e\vec{\textbf{E}}$. Here, $-\textbf{d}$ that appears in the second equation of motion arises as a result of the opposite directions of $F^+$ and $F^-$. Now, to obtain the total force, $F$, that causes the displacement, $\textbf{d}$, one needs to write $F = F^+ - F^-$. After taking $1/M = 1/M^+ + 1/M^-$, using Eq.~(\ref{eq:7}) for the electric field, and using equations similar to Eqs.~(\ref{eq:5}) and~(\ref{eq:6}) for $G$, one can arrive at

\begin {eqnarray}
\alpha_{d} = \frac{e^2}{M}\bigg[\frac{\exp[\lambda(E_F^0 - \xi)]}{(\omega_{\rm{ph}}^2 - \omega^2)}\bigg], \label{eq:19}
\end {eqnarray}  
  
where $\omega_{\rm{ph}}$ is the phonon frequency of undeformable ions. Note here that we have used the definition of the atomic polarizability given in Eq.~(\ref{eq:11}) to derive Eq.~(\ref{eq:19}). Note also that Eq.~(\ref{eq:19}) gives the displacement polarizability for deformable ions. For example, if $\lim_{\xi\rightarrow \infty}\exp[\lambda(E_F^0 - \xi)] = 0$, then the ions are infinitely rigid, and $\alpha_d \rightarrow 0$. On the other hand, $\lim_{\xi\rightarrow E_F^0}\exp[\lambda(E_F^0 - \xi)] = 1$ implies undeformable ions, because the ion radii are constant due to constant $E_F^0$. Further details can be found in Ref.~\cite{arul1}. In other words, the ion deformability in the presence of the screened electrons has been taken into account via the interaction potential constant, $G$, which is a function of the ionization energy ($\xi$). Therefore, one does not need to use the common approach of adding Eqs.~(\ref{eq:19}) and~(\ref{eq:11}) for both positive and negative ions to obtain the total polarizability~\cite{ash}. Furthermore, one can include the polarizability of the core electrons by calculating $\xi$ accurately for those core electrons of each ion in a given compound. 

Calculations on the atomic polarizability have been carried out by Woods et al.~\cite{wood} for alkali halide crystals using a different approach known as the Shell Model theory with appropriately introduced different types of short-range interaction potential constants, $\Phi_{xy}$. A similar approach (via elastic constants) was also used by Upadhyaya et al.~\cite{upa} to analyze carbides such as, TiC, ZrC and HfC. They found that the three-body interaction and free-carrier doping are also important to understand the lattice vibrations. In our theory however, $G$ takes care of the evolution of the lattice displacement and/or vibration for different elemental composition via the exponential term. The reason to use this strategy with ionization energy theory is to assist the experimenters to calculate and estimate accurately the evolution of the displacement polarizability in their nanoscale synthesis of multi-element QDs. This is important since it gives prior knowledge before the experiments are carried out, of which, details on this issue will be discussed in the next section. Secondly, this strategy avoids using the time-consuming computational method such as those based on various approximations of the density functional theory~\cite{mxu}. In summary, in Ref.~\cite{ash}, the atomic polarizability (static and isolated atoms or ions) only allows the free-electrons to be displaced. Whereas, the displacement polarizability considers the electrons as static, and only the ions are allowed to be displaced. That is why these ions are called undeformable. However, in this paper, (i) we transformed the electrons in atomic polarizability to be non free-electrons, as it should be, and (ii) using the transformation procedure given in (i) we transformed the displacement polarizability to obtain the ionic polarizability, which also includes the electronic polarizability of non free-electrons. Therefore, the ionic polarizability discussed here is for deformable ions.

\section{Discussion and analysis}

Having derived all the relevant equations, we can now try to understand the microscopic mechanisms involved in the formation of QDs in the presence of applied electric field and temperature for different QDs and elemental composition. Using the macroscopic formulation from Refs.~\cite{os1,os3,os4}, we can write the dimensionless energy of an adatom as~\cite{os1}

\begin {eqnarray}
\bar{\epsilon}_e = \frac{\lambda_a}{k_B T}\frac{\partial \vec{\textbf{E}}_{\rm{app}}}{\partial r}[p + \alpha \vec{\textbf{E}}_{\rm{app}}] = \frac{\epsilon_e}{k_BT}, \label{eq:20}
\end {eqnarray}  
  
which can be related to the surface diffusion coefficient~\cite{os1}, or also known as the surface diffusivity~\cite{sto},

\begin {eqnarray}
D_S = \lambda_a^2\frac{\omega_{\rm{ph}}}{2\pi}\exp\bigg[\frac{\epsilon_e-\epsilon_d}{k_B T}\bigg], \label{eq:21}
\end {eqnarray}  

where $\omega_{\rm{ph}} = 2k_BT/\hbar$, denotes the lattice oscillation frequency, $\hbar = h/2\pi$, $h$ is the Planck's constant, $r$ is the QD radial coordinate and $\epsilon_d$ is the surface diffusion activation energy. Equations~(\ref{eq:20}) and~(\ref{eq:21}) imply that when the applied electric field gradient, $\partial \vec{\textbf{E}}_{\rm{app}}/\partial r$ is large, then $\epsilon_e$ and $D_S$ are also large. Any increment in dimensionless energy, $\bar{\epsilon}_e$ implies an increased surface diffusivity, $D_S$ because large $\epsilon_e$ reduces the effect of $\epsilon_d$ as follows from Eq.~(\ref{eq:21}). In other words, any increment in $\epsilon_e$ will lead to the decreased surface diffusion activation energy ($\epsilon_e-\epsilon_d$) and higher rates of adatoms leaving the smaller QDs that will be available to form new QDs~\cite{os1}. Now, it is possible to incorporate the previous microscopic results given in Eq.~(\ref{eq:19}) into Eq.~(\ref{eq:20}) to obtain 

\begin {eqnarray}
\bar{\epsilon}_e = \frac{\lambda_ae^2}{k_B TM}\bigg[\frac{e^{\lambda(E_F^0 - \xi)}}{\omega_{\rm{ph}}^2}\bigg]\frac{\partial \vec{\textbf{E}}_{\rm{app}}}{\partial r}[\vec{\textbf{E}} + \vec{\textbf{E}}_{\rm{app}}]. \label{eq:22}
\end {eqnarray}  

The displacement polarizability (Eq.~(\ref{eq:19})) is used because it captures accurately the energetics required for an adatom to leave a particular surface. Whereas, Eq.~(\ref{eq:11}) deals only with electrons moving away from their ionic core. Let us now describe three physical mechanisms based on Eq.~(\ref{eq:22}). For example, to grow SiC QDs on a Si substrate, any changes to the applied electric field and surface temperature will also change the adatom (Si and C) diffusivity accordingly, as was simulated previously~\cite{os1,ho}. The other two mechanisms are related to the systematic changes in elemental composition either due to (i) size or (ii) for an entirely new QD material (for example, Ge QDs on Si). The point (i) arises from the known simulation results where the elemental composition of Si$_{1-x}$C$_x$ changes during QD growth by changing the influx ratio of Si and C~\cite{rider}. On the other hand, point (ii) is obvious if the QDs are made of different elements. For these latter two points, one can fix the surface temperature and the applied electric field, and any systematic changes in the elemental composition affect the dimensionless energy of the adatoms due to the variation of the ionization energy (see Eq.~(\ref{eq:22})). As a result, the adatom diffusivity also fluctuates accordingly. Simply put, if one considers the Si$_{1-x}$C$_x$ QDs, then the average ionization energy value for Si and C can be calculated as, $\xi_{\rm{Si}^{4+}} = 2488$ kJ mol$^{-1}$ and $\xi_{\rm{C}^{4+}} = 3571$ kJ mol$^{-1}$, where the ionization energy values prior to averaging have been obtained from Ref.~\cite{web}. Apparently, $\xi_{\rm{Si}^{4+}} < \xi_{\rm{C}^{4+}}$, therefore from Eqs.~(\ref{eq:21}) and~(\ref{eq:22}) one has a higher diffusivity for Si compared to C, i.e., $D_S^{\rm{Si}^{4+}} > D_S^{\rm{C}^{4+}}$. 

\begin{figure}[htb!]
\begin{center}
%\begin{figure}[tb!]
%\centering
\scalebox{0.3}{\includegraphics{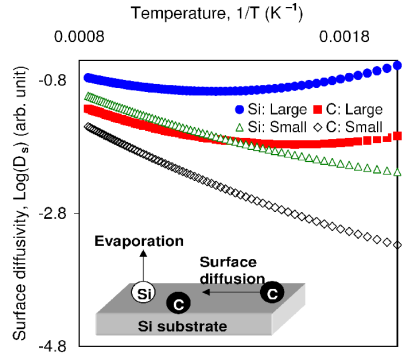}}
\caption{Arrhenius plots based on Eqs.~(\ref{eq:21}) and~(\ref{eq:22}) are given for both Si and C adatoms, in the presence of small and large $\vec{\textbf{E}}_{\rm{app}}$. INSET: Schematic diagram of a Si substrate prior to QD formation, with Si and C adatoms, and the possible evaporation and diffusion processes.}
\label{fig:3}
\end{center}
%\vspace{0.7cm}
\end{figure}

This is a telling sign that at a given temperature and/or applied electric field, one will always have higher rates of evaporation and diffusion for Si ions compared to C, from the Si substrate. Note here that the Si and C elements need not be ions, they can be atomic, as well as ions with valence states ranging from 1+ to 4+. Thus, one needs to control the influx of the Si atoms/ions (we will simply write adatom for convenience) more rigorously with growth time as compared to carbon species since the latter adatoms (carbon) have lower diffusivity and rates of evaporation from the surface of a given substrate. Focusing on the surface diffusivity of C and Si adatoms (prior to QD formation), we have plotted the $D_s$ for both Si and C adatoms in Fig.~\ref{fig:3} as a function of temperature and with small and large  $\vec{\textbf{E}}_{\rm{app}}$. The $\epsilon^{\rm{Si,C}}_d$ can be predicted from $\epsilon^{\rm{Si}}_d \propto \epsilon_{\rm{Si-Si}}$ and $\epsilon^{\rm{C}}_d \propto \epsilon_{\rm{Si-C}}$, where $\epsilon_{\rm{Si-Si}}$ (327 kJmol$^{-1}$) and $\epsilon_{\rm{Si-C}}$ (452 kJmol$^{-1}$) are the Si-Si and Si-C diatomic bonding energies, respectively, and the values are obtained from Ref.~\cite{web}. The inset of Fig.~\ref{fig:3} schematically shows two common elementary processes that occur on the Si substrate prior to QD formation, namely, the evaporation and the surface diffusion of the Si and C adatoms.

%%%%%%%%%%%%%%%%%%%%%%%%%%%%%%%%%%%%%%%%%%%%%%%%%%%%%%%%%%%%%%%%%%%%%%%%%%%%%%%%%%%%%%%%%%%%%%%%%%%%%%%%%%%%%%%%%%%%%%%%%%%%%
%%%%%%%%%%%%%%%%%%%%%%%%%%%%%%%%%%%%%%%%%%%%%%%%%%%%%%%%%%%%%%%%%%%%%%%%%%%%%%%%%%%%%%%%%%%%%%%%%%%%%%%%%%%%%%%%%%%%%%%%%%%%

The Arrhenius plots in Fig.~\ref{fig:3} are calculated from Eqs.~(\ref{eq:21}) and~(\ref{eq:22}), for both Si and C adatoms. The labels (Si,C): Small and Large in Fig.~\ref{fig:3} indicate the respective small and large $\vec{\textbf{E}}_{\rm{app}}$. This is to impose small and large contributions from $\epsilon_e$, respectively, which follows from Eq.~(\ref{eq:22}). It is clear from these Arrhenius plots that when the $\epsilon_e$ is small (small $\vec{\textbf{E}}_{\rm{app}}$), the surface diffusivity for both C and Si adatoms are as expected, which are strongly proportional to temperature. For large $\epsilon_e$, we have an interesting effect where the surface diffusivity can be made large at a lower temperature, as is shown for both Si: Large and C: Large in Fig.~\ref{fig:3}. 

This low-temperature$-$large surface diffusivity effect is entirely due to the electric dipole moment of adatoms in the presence of large $\vec{\textbf{E}}_{\rm{app}}$ (see Eq.~(\ref{eq:11})). For a larger $\vec{\textbf{E}}_{\rm{app}}$, the electronic polarization of adatoms and Si substrate electrons are larger, which give rise to stronger repulsion between the polarized electrons from the Si substrate and the adatoms, which in turn lead to larger surface diffusivities. The second issue from Fig.~\ref{fig:3} is the fact that the respective surface diffusivity of Si adatoms, for both small and large $\vec{\textbf{E}}_{\rm{app}}$, is larger than the diffusivity of C adatoms. The reason for this different-element effect originates from both $\epsilon_e$ and $\epsilon_d$. Since $\xi_{\rm{Si}} < \xi_{\rm{C}}$ we obtain $\epsilon_e^{\rm{Si}} > \epsilon_e^{\rm{C}}$, and using $\epsilon^{\rm{Si}}_d (\propto \epsilon_{\rm{Si-Si}}) < \epsilon^{\rm{C}}_d (\propto \epsilon_{\rm{Si-C}})$ that leads to $|\epsilon_e^{\rm{Si}} - \epsilon_d^{\rm{Si}}|$ $<$ $|\epsilon_e^{\rm{C}} - \epsilon_d^{\rm{C}}|$, which explains why the Si plots in Fig.~\ref{fig:3} are always above the C plots for a given $\vec{\textbf{E}}_{\rm{app}}$. Note here that $\epsilon_d^{\rm{Si}} > \epsilon_e^{\rm{Si}}$ and $\epsilon_d^{\rm{C}} > \epsilon_e^{\rm{C}}$. On the other hand, in the absence of $\vec{\textbf{E}}_{\rm{app}}$ and/or at $T = 0$, the inequality, $|\epsilon_e^{\rm{Si}} - \epsilon_d^{\rm{Si}}|$ $<$ $|\epsilon_e^{\rm{C}} - \epsilon_d^{\rm{C}}|$ becomes $\epsilon_d^{\rm{Si}}$ $<$ $\epsilon_d^{\rm{C}}$ or $\epsilon_d^{\rm{Si}}$ $\geq$ $\epsilon_d^{\rm{C}}$, which is entirely dependent on the $\epsilon_d$ because $\epsilon_e \rightarrow 0$. Note here that $\epsilon_d$ is a constant for a given element and does not vary in the presence of $\vec{\textbf{E}}_{\rm{app}}$ and at a finite temperature. On the other hand, $\epsilon_e$ has been allowed to vary with respect to $\vec{\textbf{E}}_{\rm{app}}$, $T$ and $\xi$. 

Therefore, $D_S^{\rm{Si}} > D_S^{\rm{C}}$, for a Si substrate and for given $\vec{\textbf{E}}_{\rm{app}}$, $T$ and $\xi$, which implies that Si and C incoming fluxes ([Si] and [C], respectively) need to be controlled in such a way that [Si]/[C] $>$ 1. If the average influx ratio, [Si]/[C] $\leq$ 1, then the QDs will always have C cores. In the other extreme, Si-core QDs can only be obtained if [Si]/[C] $\gg$ 1, for given $\vec{\textbf{E}}_{\rm{app}}$ and $T$. One can thus use the ionization energy theory to narrow down the influx ratio for any multi-element QDs. For example, to grow Si$_{1-x}$C$_x$ QDs, with a large carbon content ($x > 0.5$) is relatively easy by choosing the average ratio of [Si]/[C] $\leq$ 1. On the other hand, to obtain $x \leq 0.5$, one needs the average ratio of [Si]/[C] $>$ 1. Subsequently, we can extend this strategy to calculate $D_S$ for an adatom to leave their QDs, and back to the Si substrate. Using Si$_{1-x}$C$_x$ QD system as an example, we can again use the inequalities introduced earlier to estimate $D_S^{\rm{Si,C}}$. Evidently, the ionization energy inequality remains the same where, $\xi_{\rm{Si}} < \xi_{\rm{C}} \rightarrow \epsilon_e^{\rm{Si}} > \epsilon_e^{\rm{C}}$. Whereas, the respective adatom diffusion activation energies for Si and C are given by $\epsilon^{\rm{Si}}_d \propto [(x)\epsilon_{\rm{Si-C}} + (1-x)\epsilon_{\rm{Si-Si}}]$ and $\epsilon^{\rm{C}}_d \propto [(1-x)\epsilon_{\rm{Si-C}} + (x)\epsilon_{\rm{C-C}}]$, where $\epsilon_{\rm{C-C}}$ = 607 kJmol$^{-1}$. Hence, following the above analysis, we will eventually obtain the same result discussed earlier, which is $|\epsilon_e^{\rm{Si}} - \epsilon_d^{\rm{Si}}|$ $<$ $|\epsilon_e^{\rm{C}} - \epsilon_d^{\rm{C}}|$ for all $x$. As a consequence, regardless of the QDs atomic composition ($x$), Si adatoms will always have a strong tendency to leave the QDs. This tendency becomes the strongest for $x = 1$, while the weakest effect occurs when $x = 0$. On the contrary, C adatoms are always favorable to form QDs on Si substrate, as compared to Si adatoms, for all $x$ and for given $\vec{\textbf{E}}_{\rm{app}}$ and $T$.      

Finally, it is now possible to discuss the experimental observations on the growth of SiC thin films (there are no available experimental result that relates SiC QDs and $x$), either on Si or SiC substrates. These two substrates will give similar results on the influx concentration ratio because $D_S^{\rm{Si}} > D_S^{\rm{C}}$ is always true for both Si or SiC substrates, as predicted above. However, the experimental observations~\cite{geo,moto,moto2,liu,qj,qj2} require the influx ratio [Si]/[C] $<$ 1 to obtain stoichiometric SiC. This result, when looked closely, indeed agrees with the analysis presented above. Firstly, the theoretical results highlighted above assumed the surface Si-atoms from the substrate do not act as adatoms to form QDs. Therefore, any Si-adatom contributed by the Si substrate needs to be incorporated into the [Si] as [Si]$_{\rm{INF}}$+[Si]$_{\rm{SUB}}$, where [Si]$_{\rm{INF}}$ and [Si]$_{\rm{SUB}}$ denote the concentration of Si adatoms from the influx and the substrate, respectively. For example, in the presence of Si adatoms from the substrate, the correct ratio should be written as ([Si]$_{\rm{INF}}$+[Si]$_{\rm{SUB}}$)/[C] $>$ 1, which is as predicted. If we ignore [Si]$_{\rm{SUB}}$, then we obtain [Si]$_{\rm{INF}}$/[C] = [Si]/[C] $<$ 1. This is the reason why in the reported growth processes, the required concentration is [Si]/[C] $<$ 1. For example, in a chemical vapor deposition process this can be considered as the concentration for the choice of precursor fluxes, e.g. [SiH$_4$]/[CH$_4$] $<$ 1. The presence of hydrogen will not affect the result $D_S^{\rm{Si}} > D_S^{\rm{C}}$ because both Si and C contain hydrogen as SiH$_4$ and CH$_4$. Moreover, this comes as no surprise due to larger surface diffusivity for Si adatoms and lower Si-Si bond strength, as compared to C adatoms. In addition, the formation of amorphous carbon and/or polycrystalline SiC thin films has been reported, even with only a small increase in the C concentration, [Si]/[C] $\approx$ 0.7 (Ref.~\cite{qj}). This finding also agrees with the conclusion stated above, which is due to the lower surface diffusivity of C adatoms. 

Further experimental evidence comes from the work of Emtsev et al.~\cite{em}, in which, the wafer-size graphene layer was produced by annealing the SiC sample. They found that Si atoms tend to leave the SiC layers under a given condition, be it in the presence of Ar gas or annealed in ultra-high vacuum. This finding is in fact agrees with the microscopic theory presented here, where the Si atoms will always have a higher tendency to diffuse and evaporate from the SiC layers, as compared to carbon. This leaves the carbon behind as graphene layers (graphitization)~\cite{em}. Add to that, our theory also suggest that the annealing process with Ar gas can be further controlled by using the applied electric field at a lower annealing temperature. 

Importantly, the ionization energy theory presented here can be used to study the diffusion of any adatoms on any non free-electron substrate materials. If the substrate is a free-electron metal, then Eq.~(\ref{eq:4}) reduces to the standard time-independent Schrodinger equation given by, $\hat{H}\varphi = E\varphi$, where one needs other theoretical and computational methods to solve it by means of variational principle~\cite{mxu}. It is also worth mentioning that when one compares the diatomic bonding energies alone [Si-C (452 kJmol$^{-1}$) $>$ Si-Si (327 kJmol$^{-1}$)] for C and Si adatoms on Si substrate, then one can surmise that indeed Si tends to diffuse more easily than C adatom. However, this comparison is only true for single-element, non free-electron substrate. For example, we cannot use it to explain the experimental results obtained by Emtsev et al.~\cite{em}, where the Si atoms (compared to C) tend to leave the SiC substrate when it is exposed to high vacuum and temperature. As such, the advantage of using both the diatomic bonding energies and the ionization energy based many-body Hamiltonian is two-fold: it gives accurate microscopic and rigorous physical explanations on the diffusivity of any adatoms on (i) any substrates, and (ii) diffusion and evaporation from any substrate. We stress here that we did not stretch the meaning of the diatomic bonding energies beyond its original definition, which is important for accurate analysis and calculations. Of course, the substrates stated in (i) and (ii) can be single- and/or multi-element compounds, apart from the fact that they must be non free-electron metals.

\section{Conclusions}

We have derived both the semiclassical and quantum mechanical version of atomic polarizability that treats all the electrons as strongly correlated with discreet energy levels. Subsequently, the derivation for the displacement polarizability as a function of the ionization energy enabled us to avoid the undeformable-ion formalism. Consequently, we have obtained the total polarizability in order to evaluate the effect of the surface diffusion of different adatoms on a given substrate. We have further found that by knowing how the ionization energy changes for different elemental composition, one can fine-tune the incoming fluxes of precursor species for a given surface temperature and the applied electric field to grow the required QDs, with a particular composition. The incoming fluxes should be controlled because the adatoms with a lower ionization energy have a strong tendency to diffuse on the surface. Finally, using the results of our work, it is possible to engineer the growth of multi-element quantum dots on any plasma-exposed surfaces (non free-electron metals).    

\section*{Acknowledgments}

A.D.A. would like to thank the School of Physics, University of Sydney for the USIRS award, and Kithriammah Soosay for the partial financial support. K.O. acknowledges the partial support from the Australian Research Council (ARC) and the CSIRO.

\end{document}